\documentclass[superscriptaddress,twocolumn,showpacs,prl,amsmath,amssymb]{revtex4-1}

\usepackage{jhwmath}

\begin{document}

\title{Breakdown of the coherent state path integral: \\ Two simple
  examples}
\date{\today}
\author{Justin H.\ \surname{Wilson}}
\affiliation{Joint Quantum Institute and Condensed Matter Theory
  Center, Department of Physics, University of Maryland, College Park,
  Maryland 20742-4111, USA} 
\author{Victor \surname{Galitski}}
\affiliation{Joint Quantum Institute and Condensed Matter Theory
  Center, Department of Physics, University of Maryland, College Park,
  Maryland 20742-4111, USA} 

\begin{abstract}
We show how the time-continuous coherent state path integral
breaks down for both the single-site Bose-Hubbard model and the
spin path integral. Specifically, when the Hamiltonian is
quadratic in a generator of the algebra used to construct
coherent states, the path integral fails to produce correct
results following from an operator approach. As suggested by
previous authors, we note that the problems do not arise in the
time-discretized version of the path integral. 
\end{abstract}

\maketitle

Path integrals are widely known for being an alternate formulation of
quantum mechanics, and appear in many textbooks as a useful
calculational tool for various quantum and statistical mechanical problems (e.g.,
\cite{altland_condensed_2010,wen_quantum_2007,
  peskin_introduction_1995}). From their inception,  
there has been the problem of writing down a path integral for any
system that can be described by a Hilbert space equipped with a
Hamiltonian. One way to approach this problem is with what is known
now as the generalized coherent state path integral
\cite{papadopoulos_continuous_1977, 
berezin_feynman_1980} which generalizes the coherent state path integral
for a harmonic oscillator. The key observation with path integration
\cite{wen_quantum_2007} 
is that, given a Hamiltonian $H$, the propagator, $e^{-it H}$, at
some time $t$ can be broken up into 
$N$ slices, $(e^{-itH/N})^N$, and in between each multiplicative term
one inserts an (over-)complete set of states parametrized by a
continuous parameter; generalized coherent states meet that
criteria. If we take
$N\rightarrow\infty$ we get the time-continuous formulation. Coherent
state path integrals have become widely and routinely used in many 
areas of physics (see the many papers collected in
\cite{klauder_coherent_1985}).

Glauber coherent states \cite{glauber_coherent_1963} are usually
understood as the most classical states 
associated with the harmonic oscillator. They obey the
classical equations of motion for a harmonic oscillator and are
minimal uncertainty states. Perelomov and Gilmore
\cite{perelomov_coherent_1972,gilmore_geometry_1972} extended the
definition of coherent  
states to Lie algebras other than the Heisenberg algebra (i.e., the harmonic
oscillator algebra). Since then, these ``generalized'' coherent states
have been used in a number of applications (see 
\cite{zhang_coherent_1990, perelomov_generalized_1986} for more on
them). In particular, the 
coherent states form an overcomplete basis (with a continuous
label) and therefore admit a resolution of the identity, which is a
necessary ingredient for the construction of a path integral.  For the
harmonic oscillator, coherent states are
represented by a complex number $\alpha$, but for coherent states constructed
with $\mathfrak {su}(2)$ (spin), they are points
on the Bloch sphere, $S^2$. 

For the case of the harmonic oscillator, it is commonly
known that one can easily go between the
normal-ordered Hamiltonian (all
annihilation operators commuted to the 
right) and the coherent state path integral
\cite{altland_condensed_2010}; this is due to the fact 
that coherent states are eigenvectors of the annihilation
operator. For the general coherent state path
integral, the ``classical'' Hamiltonian  
in the path integral is just the expectation value of the quantum
Hamiltonian with a coherent state. This prescription results in
some notable exactly solvable cases, but all such cases involve
\emph{non-interacting} terms which are essentially linear in the
algebra generators used to construct the coherent-states. When the
Hamiltonian involves terms that are
non-linear in generators (interactions), this prescription seems to
fail, as this letter demonstrates.

Path integrals are widely used for developing perturbative expansions
in terms of Feynman diagrams, for non-perturbative techniques (e.g.,
the instanton method), and for deriving effective theories
\cite{peskin_introduction_1995,wen_quantum_2007}, but despite the many
successes of path integrals, they have been on very shaky 
mathematical grounds (for a small ``slice'' of this history, see
\cite{klauder_feynman_2003}). In particular, the spin coherent state path
integral has sometimes produced
(quantitatively) incorrect results \cite{enz_spin_1986, funahashi_coherent_1995,
  funahashi_more_1995, belinicher_instanton_1997, shibata_note_1999}
unless the time-discretized version is employed
\cite{solari_semiclassical_1987, belinicher_instanton_1997}. These
problems with the time-continuous path integral were mostly solved by
Stone \emph{et al.}\ 
\cite{stone_semiclassical_2000} by identifying an anomaly in the
fluctuation determinant which added an extra phase to the
semi-classical propagator. Kochetov had also found this phase in a
general context \cite{kochetov_quasiclassical_1998}. Furthermore,
Pletyukhov~\cite{pletyukhov_extra_2004} related the extra phase in the
spin path integral back to
Weyl ordering the Hamiltonian in the
case of the harmonic oscillator (in the simplest case, Weyl ordering
corresponds to symmetrically 
ordering annihilation and creation operators). Additionally, Weyl
ordering has been considered in the Bose-Hubbard case in
\cite{polkovnikov_phase_2010}. Unfortunately, this solution does not 
explain the present breakdown under consideration. 

In this letter, we outline another problem with the time-continuous
coherent state path integral. This problem manifests itself in two
simple examples: (i) the spin-coherent state path integral and (ii)
the harmonic oscillator coherent state path integral (in 
particular, the single-site Bose-Hubbard model). The single-site
Bose-Hubbard Hamiltonian is a minimal model that demonstrates the
problem with the normal-ordered path integral. However, the problem
itself is more general than the toy model considered here and clearly
persists in more complicated models, including lattice Bose-Hubbard
models. We use an exact method of calculating the partition function,
first used for spin (with $H=S_z$) by Cabra \emph{et
  al.}~\cite{cabra_path_1997}, and demonstrate that the exact result
differs from the correct partition function in the cases of both
normal-ordering of operators (as prescribed by most textbooks) and
when using Weyl ordering (i.e., it cannot be accounted for with the phase
anomaly found by Solari and Kochetov
\cite{solari_semiclassical_1987,kochetov_quasiclassical_1998} and
elaborated on by Stone \emph{et al.}~\cite{stone_semiclassical_2000}).

We begin with the coherent state path integral for spin with the
standard SU(2) algebra defined on the operators $\{S_x, S_y, S_z\}$
with $[S_i, S_j] = i \epsilon_{ijk} S_k$,
and we define our Hilbert space by taking the matrix representation of
the SU(2) group in $(2s+1)$-by-$(2s+1)$ matrices ($s$ being the spin
of the system). Irrespective of the
algebra, we can in general define a Hermitian matrix $H$ that acts on
states in our Hilbert space, and this will be our
Hamiltonian. Usually, $H$ is polynomial of algebra generators.

If $\ket s$ is the maximal state of $S_z$ in our spin-$s$ system, then we can define
spin-coherent states as $\ket{\bvec n} = e^{-i\phi S_z} e^{-i\theta S_y} \ket s$ 
where $(\theta,\phi)$ are coordinates on the sphere $S^2$ along the
unit vector $\bvec n$ (i.e., a point on the standard Bloch
sphere). These coherent states are  
overcomplete such that $ \frac{2s+1}{4\pi} \int_{S^2} \rd \bvec n \ket{\bvec n}
  \bra{\bvec n} = 1$ 
where $\rd\bvec n = \rd \phi \,\rd(\cos\theta)$ is the standard measure
on $S^2$.  Using this continuous, overcomplete basis, one can derive
the standard path integral for the partition function for spin from
$\mathcal Z = \tr e^{-\beta H}$ in the standard way \cite{altland_condensed_2010} discussed in the introduction:
\begin{multline}
 \mathcal Z' = \int \calD \bvec n(\tau) \,
  \exp\inbr{-\int_0^\beta \rd \tau \inbk{ -\braket{\bvec n(\tau) |
        \pd_\tau  \bvec n(\tau)}\right.\inlbreak\left. + \braket{\bvec n(\tau)|H|\bvec n(\tau)} }}\label{eq:1}.
\end{multline}
We call the partition function as given by the
time-continuous path integral $\mathcal Z'$ in order to
distinguish it from $\mathcal Z = \tr e^{-\beta H}$ since we will find
that in general they may not agree.
The path integral is over all closed paths (since it is the
parititon function). The first term in the action for
Eq.~(\ref{eq:1}), $\braket{\bvec n| \pd_\tau \bvec n}$, is the Berry
phase term and in $(\theta,\phi)$ coordinates $-\braket{\bvec n |
  \pd_\tau \bvec n} = -si (1 - \cos \theta)\pd_\tau \phi$. 

In order to highlight what is wrong with a na\"ive use of
Eq.~(\ref{eq:1}), we employ the method of Cabra \emph{et al.}\
\cite{cabra_path_1997}. We assume 
$\braket{\bvec n(\tau) | H | \bvec n(\tau)} = 
H(\cos\theta(\tau))$ for some function $H(x)$ (this is true if and
only if $H$ is diagonal). This puts the $\phi$ dependence of the
action solely in the Berry phase term of the action. We then integrate
the Berry phase term by parts; the boundary term is just $\Delta
\phi (1 - \cos\theta(0))$ with $\Delta \phi = \phi(\beta)  - \phi(0)
= 2 \pi k$ for any integer $k$ and $\cos\theta(\beta)=\cos\theta(0)$
since our paths are closed. We must sum over the different topological
sectors defined by the integer $k$ (i.e., how many times $\phi$ wraps
around the sphere). Thus, our only $\phi$ dependence is multiplying
$\rdf{\cos\theta} \tau$ from integrating by parts, and we use standard
identity  for functional integrals
$  \int \calD \phi \, e^{-i \int_0^\beta \rd \tau \, \phi(\tau)
    f(\tau)} = \delta(f)$, 
to get that $\cos\theta$ must be constant (i.e.,
$\rdf{\cos\theta}{\tau} = 0$). This $\delta$-function 
allows us to do the path
integral over $\calD(\cos \theta)$, except for the initial value which
we call $x := \cos\theta(0)$.  Taking all of this into account, the
path integral can then be written as 
\begin{align}
\mathcal Z' = \sum_{k=-\infty}^\infty\int_{-1}^1 \rd x \, e^{2\pi i k
   s(1-x) - \beta H(x)}.\label{eq:2}
\end{align}
The sum over $k$ can be evaluated as a sum of delta functions of the form
$\delta(s(1-x)-n)$ for all integers $n$. Since $x$ is in the interval
$-1$ to $+1$, only finitely many $n$ contribute ($n=0$ to $n=2s$ to be
exact). We can rewrite the sum over $n$ as a sum
over $m:=s-n$ and we get the answer (dropping overall constants)
\begin{align}
  \mathcal Z' = \sum_{m=-s}^s e^{-\beta H(m/s)}.\label{eq:5}
\end{align}
Eq.~(\ref{eq:5}) looks very promising, but $H(m/s)$ is not the same as
$\braket{m|H|m}$. First let us see where it \emph{does} work. Take the simple
Hamiltonian $H=S_z$, then $\braket{\bvec n| H |\bvec n} = s \cos
\theta$, and thus $H(x) = s x$. This immediately yields
\begin{align}
  \mathcal Z'_{H=S_z} = \sum_{m=-s}^s e^{-\beta m},
\end{align}
and it is easily calculated (in operator language) that $ \mathcal
Z'_{H=S_z} = \mathcal Z_{H=S_z}$. The two methods agree for the
particular Hamiltonian $H=S_z$ (the case considered by Cabra \emph{et
  al.}\ \cite{cabra_path_1997}). On the other hand, if we take
$H=S_z^2$ and $s=1$, we can evaluate 
$\braket{\bvec n | S_z^2|\bvec n} = \frac12\inp{
  \cos^2\theta +1}$; from which we have
\begin{align}
  H(x) = \frac12\inp{ x^2+ 1}.\label{eq:4}
\end{align}
Thus, $\mathcal Z'_{H=S_z^2}= 2 e^{-\beta} + e^{-\beta/2}$, but this conflicts with
${\mathcal Z}_{H=S_z^2} = 2 e^{-\beta} +1$ 
by more than just a multiplicative constant. Thus, we have
$\mathcal Z'_{H=S_z^2} \neq \mathcal Z_{H=S_z^2}$ for $s=1$,
and in fact $\mathcal Z'_{H=S_z^2} \neq \mathcal Z_{H=S_z^2}$
for all $s>1/2$.

Importantly, the two methods agree for any Hamiltonian when
$s=1/2$. This comes 
from the fact that any (diagonalized) Hamiltonian for a two state
sytem ($s=1/2$) can be 
written as $H = a + b S_z$ (in fact $H=S_z^2 = 1/4$), and the above
method gives $\mathcal Z'  = \mathcal Z$ when $H=a+b S_z$.

Also, if we let our Hamiltonian be $H= S_z^2/ s^2$, then in the limit of
$s\gg 1$ Eq.~(\ref{eq:5}) reproduces the correct result. This is a
general result for Hamiltonians that are finite polynomials of 
$S_z/s$, and suggests that ``semiclassically'' (i.e. $s$ tends
to infinity), we will still arrive at sensible results.

Agreement can also be forced by considering $H(x)=x^2$ instead of
Eq.~(\ref{eq:4}), but this corresponds to replacing $S_z$ with
$\braket{S_z}$ in the Hamiltoinian \emph{instead} of just considering
$\braket{H}$. In the $H=S_z^2$ case, it is the difference between
considering $\braket{S_z^2}$ and $\braket{S_z}^2$; the latter gives
correct results. There is no \emph{a priori} reason for this
construction.

\newcommand{\vspan}{\operatorname{{\mathrm span}}}
To motivate looking for this same issue in a system with the
Weyl-Heisenberg algebra (i.e., the harmonic oscillator algebra),
it is known \cite{gilmore_lie_2008} that one can contract
$\mathfrak u(2)$ (since we constructed our coherent states 
for spins with $\mathfrak{su}(2)$) into the Weyl-Heisenberg
algebra by considering
$  \mathfrak u(2) = \vspan\{ S_0, S_x, S_y, S_z\} = \mathfrak u(1)
  \oplus \mathfrak{su}(2),$
where we define $[S_0,S_i]=0$. Then define the operators $J_0:=S_0$,
$J_{1,2}(\epsilon):=\epsilon S_{y,x}$, and $J_3(\epsilon) := S_0 +
\epsilon^{-2}S_z$ to get the commutation relations
 $ [J_3, J_{1,2}]  = \mp i J_{2,1}$, $[J_1,J_2] = -i \epsilon^2 J_3 +
  i J_0$, and $[J_0, J_i]  = 0$.  If we let $\epsilon\rightarrow 0$,
  we recover exactly the Weyl-Heisenberg algebra: 
$ \mathfrak h_4 = \vspan\{ 1, x, p , a\hc a\}$  
with $[x,p]=i$, $[a\hc a, x] = -ip$, $[a\hc a, p]  = i x$.
Observe that $S_z^2$ is related to $a\hc a$ in this
contraction, so we might suspect that terms quadratic in $a\hc a$
might give problems like those found with the spin-coherent state path integral.

A Hamiltonian that uses the Weyl-Heisenberg
algebra to construct its coherent states is the Bose-Hubbard
model. For a single site, we can write
\begin{align}
  H = - \mu n + \frac U 2 n (n - 1),\label{eq:3}
\end{align}
where $n= a\hc a$ and the $a$ ($a\hc$) is the annihilation (creation)
operator for the algebra $[a,a\hc]=1$. The form $n(n-1)=a\hc a\hc a a$
comes from the normal ordering 
required from a path integral of the form
\begin{multline}
  \mathcal  Z' = \int \calD^2 z \,\exp\inbr{-\int_0^\beta \rd \tau
  \inbk{\frac 1 2 ( z^* \dot z - \dot z^* z) \right.\inlbreak\left. - \mu |z|^2 + \frac U 2 |z|^4}}.
\end{multline}
We can solve this path integral with the same method used to obtain
Eq.~\eqref{eq:5} in the spin-coherent state path integral. Let $z
= \sqrt{n} e^{i\theta}$, so that the measure becomes $\calD^2z = \calD n
\calD \theta$ and the action becomes $S =\int \rd \tau (i n \dot \theta  - \mu n +
\frac U 2 n^2)$. 
Integrating by parts on the $ n \dot \theta$ term then integrating over $\calD
\theta$ will fix $n$ to be  
constant, and the boundary term will fix $n$ to be an
integer. Since $n$ is radial, it can only be positive so we directly obtain
\begin{align}
  \mathcal Z' = \sum_{n=0}^\infty e^{\mu  n\beta - \frac U 2 n^2\beta}.\label{eq:8}
\end{align}
But this differs from the partition function that we can easily
calculate in operator language:
\begin{align}
  \mathcal Z = \sum_{n=0}^\infty e^{\mu n\beta - \frac U 2 n(n-1)\beta}.
\end{align}
We see that a similar problem to that of the spin coherent
state path integral here. To see it explicitly, for $U\gg 1$,
we have $\mathcal Z' \sim 1 + e^{\mu - U/2} + \cdots$, but
$\mathcal Z \sim 1+ e^{\mu} + e^{2\mu - U} + \cdots$. With
different asymptotics, $\mathcal Z$ and $\mathcal Z'$ are
different expressions. Note that if we let $\mu\rightarrow \mu + \frac
U 2$ in $\mathcal Z'$, that we will get same result. This substitution
for $\mu$ corresponds to 
replacing $n$ in Eq.~(\ref{eq:3}) by $\braket{n}=|z|^2$ when
writing down our action (so instead of $\braket{n^2}$, one gets
$\braket{n}^2$). 




We now compare this to the semiclassical result. Still considering
Eq.~(\ref{eq:3}), let change our algebra slightly to incorporate a 
small parameter (akin to the standard $\hbar\rightarrow 0$ for
normal semiclassics): $h^{-1}$, the representation index; see $\gamma$
defined in  \cite{kochetov_quasiclassical_1998}. We note here that
different $h$'s change the coherent states $\ket z$ in the following way: if
$z=\frac1{\sqrt2}(x+iy)$, then $x=q/c$, $y=p/d$, and $h =
\hbar/(cd)$ (and $[a,a\hc] = h$). We have used $q$ and $p$ as the
standard position and momentum for the harmonic oscillator. Up until
now we have been considering $h=1$. 

We can write the propagator
between two coherent states $\ket{z_i}$ and $\ket{z_f}$ using a
Hubbard-Stratonovich transformation and the propagator for the
harmonic oscillator:
\begin{align}
  K(z_f^*,z_i;t) & = \braket{z_f| e^{-iH T/h}|z_i} \\
 & = \sqrt{\frac{i T}{2\pi U h}} \int \rd \omega  \, e^{ \frac1 h \Phi_\omega
   +\frac12 i \omega T + \frac i 8 U h T},\label{eq:7}
\end{align}
where we have defined 
\begin{align}
    \Phi_\omega &= z_f^* z_i e^{i(\omega+\mu)T} +
    \frac{iT}{2U}\omega^2 - \frac12(|z_i|^2 + |z_f|^2).
\end{align}
In path integral notation we can write out the propagator~\cite{kochetov_quasiclassical_1998}
\begin{align}
    K(z_f^*,z_i;T) = \int_{z(0)=z_I}^{z^*(T)=z^*_f} \calD^2 z \, \exp\inbr{\Phi[z,z^*]/h},
\end{align}
where $\Phi  = \Gamma + S$,
\begin{align}
  \Gamma & = \frac12\inbk{ z^*_f z(T) +
  z^*(0) z_I - |z_f|^2 - |z_I|^2} ,\\
S & =  \frac12\int_0^T \rd t
\, (z \dot z^* - z^* \dot z) - i  \int_0^T \rd t\, \braket{z|H|z}.
\end{align}

Performing the standard semiclassical analysis and algebra (see
\cite{kochetov_quasiclassical_1998} and
\cite{stone_semiclassical_2000}) the semiclassical propagator takes
the form
\begin{multline}
  K_{\text{sc}}(z_f^*,z_i;T) = \sum_\omega \inp{\frac{iT}{hU}}^{1/2}
  \inp{\frac1h\pdf[2]{\Phi_\omega}{\omega}}^{-1/2}  \\ \times \exp\inbk{\frac1h
    \Phi_\omega +\frac i2(\omega+\mu)T-i\Delta},\label{eq:6} 
\end{multline}
where the sum is over solutions to the consistency equation given by
$\pdf{\Phi_\omega}\omega = 0$ or $\omega = - U z_f^* z_i e^{i(\omega+\mu)T}$,
and we have defined $\Delta  =  \frac 1 2 (\mu + 2\omega) T$. 
The term $\Delta$ comes from the fixing of the fluctuation determinant anomaly
described in detail by Stone \emph{et al.}\
\cite{stone_semiclassical_2000} for the SU(2) case. However, if we
try to get Eq.~(\ref{eq:6}) by using the method of steepest descent on
Eq.~(\ref{eq:7}) with $h\rightarrow 0$, we won't get the same result.
This is because of what has been shown 
by others \cite{kochetov_quasiclassical_1998, pletyukhov_extra_2004}:
that the semiclassics 
will give results consistent with the 
Weyl ordering of the Hamiltonian (na\"ively ordering all $a$ and
$a\hc$'s symmetrically). The usual normal ordered Hamiltonian takes
the form (inserting $h$'s) $H = -\mu n + \frac U 2 n (n - h)$ while
the Weyl ordered Hamiltonian takes the form (up to a constant) $H_W =
-\mu n + \frac U 2 n (n + h)$.  If we derive Eq.~(\ref{eq:7}) for
$H_W$, we will find the the steepest descent exactly agrees with
Eq.~(\ref{eq:6}) just as expected \cite{kochetov_quasiclassical_1998,
  pletyukhov_extra_2004}.  

While the semiclassical result is not a new one, it shows that the
path integral is not dealing with the same Hamiltonian. Unfortunately,
our exact calculation suggests that the path integral is dealing with
$H'=-\mu n + \frac U 2 n^2$ while semiclassics suggests it is dealing
with $H_W = - \mu n +\frac U 2 n(n+1)$. These two methods differ but
both are not the Hamiltonian under consideration. In the case
of the Weyl ordered Hamiltonian, we can write our original Hamiltonian
in Eq.~(\ref{eq:3}) as $H = H_W - U n$ which is Weyl ordered (up to a
constant). This ordering can be used to modify the path integral by an 
extra term $-Un$.  This correction to the path integral suggested by
Weyl ordering does not 
fix the exact calculation as can be easily shown, but it does
motivate an \emph{ad hoc} correction to the path integral to ``fix''
our exact calculation. We use the following action (going back to $h=1$):
\begin{align}
  S =\int \rd t \inp{ -\mu |z|^2 + \frac U 2 |z|^2(|z|^2-1)}.
\end{align}
This action is constructed by just changing the operator $n$ to a function $|z|^2$;
while this gives correct results with the method which gives Eq.~(\ref{eq:8}),
there is no \emph{a priori} reason to suspect this of being the
action. Similarly, if in the spin-coherent state path integral, we replace the operator
$S_z$ with its expectation value $\braket{\bvec n| S_z| \bvec n}$
everywhere, we will get the correct result. This means, in
particular, for $H=S_z^2$ that instead of $\braket{S_z^2}$ in the spin-path
integral we have $\braket{S_z}^2$. In general, if one substitutes the
generators of the coherent 
states in the Hamiltonian with their expectation value, one obtains the
correct result for $\mathcal Z$ with the methods used to derive
Eq.~(\ref{eq:5}) and Eq.~(\ref{eq:8}).

Corrections aside, a simple way to see what has gone wrong is to
return to Eq.~(\ref{eq:4}).
This $H(x)$ function can not achieve the value $0$, but $H=S_z^2$
clearly has such an eigenvalue. This is due to the fact that for
higher dimensional representations of SU(2) not every eigenvector of
$S_z$ can be rotated into another with a standard SU(2) rotation.
On the other hand, the coherent states we used are a complete set for even
higher dimensional representations, so in principle, we should not
lose any information about the $m=0$ state. Continuity in $\bvec n$
seems to be the culprit: $H(x)$
came from a time discretized form (between time slices $j$ and $j+1$) 
$ \braket{\bvec n_{j+1}| S_z^2 | \bvec n_j}$,
and we have $\braket{\bvec n| S_z^2 | -\bvec n}
= 0$, so $ \braket{\bvec n_{j+1}| S_z^2 | \bvec n_j}$ can attain zero,
but not for any paths that are ``close'' 
to each other (i.e. $\bvec n_j \approx \bvec n_{j+1}$) as the
continuous time path integral assumes. As such, the 
discrete time path integral (before a continuity
assumption is imposed) can unambiguously give the correct results to
a calculation. 

To conclude, in the time-continuous formulation of the path integral,
neither the action suggested by Weyl-ordering nor the action
constructed by normal ordering give the correct result when evaluating
$\mathcal Z$ via path integrals.

{\em Acknowledgements} -- We thank Michael Levin for stimulating
conversations. This research was supported by the NSF CAREER award, DMR-0847224.

\bibliography{references}

\end{document}